\documentclass[prl,superscriptaddress, preprintnumbers,twocolumn,showpacs]{revtex4}
\usepackage{amsmath}
\usepackage{amssymb}
\usepackage{amsthm}
\usepackage{amsfonts}
\usepackage{algorithmic}
\usepackage{enumerate}
\usepackage{latexsym}
\usepackage{graphicx}
\usepackage{bm}
\usepackage{subfigure}
\usepackage[dvips]{color}

\begin{document}

\title{Localization and Mobility Gap in Topological Anderson Insulator}
\author{Yan-Yang Zhang, Rui-Lin Chu, Fu-Chun Zhang and Shun-Qing Shen}
\affiliation{Department of Physics, The University of Hong Kong, Pokfulam, Hong Kong,
China}
\date{\today}

\begin{abstract}
It has been proposed that disorder may lead to a new type of
topological insulator, called topological Anderson insulator (TAI).
Here we examine the physical origin of this phenomenon. We calculate
the topological invariants and density of states of disordered model
in a super-cell of 2-dimensional HgTe/CdTe quantum well. The
topologically non-trivial phase is triggered by a band touching as
the disorder strength increases. The TAI is protected by a mobility
gap, in contrast to the band gap in conventional quantum spin Hall
systems.  The mobility gap in the TAI consists of a cluster of
non-trivial subgaps separated by almost flat and localized bands.
\end{abstract}

\maketitle

\section{I. Introduction}

Recent development in topological insulators (TI) has greatly
enhanced our understanding of topological properties in condensed
matter \cite{Hasan2010,Qi2010,Moore2010}. Band insulators with time
reversal symmetry can be classified by a $Z_{2}$ topological
invariant $\nu $ associated with the occupied bands, $\nu =0$ for
topologically trivial phase and $\nu =1$ for non-trivial phase\cite
{Kane2005A,Fu2007,Moore2007,Qi2008}. In two dimensions (2D), the TI
($\nu =1$) exhibits quantum spin Hall effect, whose edge currents
are robust against weak non-magnetic
disorder\cite{Bernevig2006,Kane2005A,Kane2005B}. This
dissipation-less transport can only be destroyed by extremely strong
disorder, which drives the system into a traditional Anderson
insulator\cite {Hatsugai1999,Li09}. The 2D TI has been
experimentally realized in HgTe/CdTe quantum wells, where the
thickness of the quantum well can be varied to tune the system
between TI and normal insulator \cite{Konig-06science,Konig2008}.

Recent numerical simulation reveals an interesting new phase, called
topological Anderson insulator (TAI)\cite{Li09}. The TAI is a
reentrant TI due to disorder: the disorder drives a 2D topologically
trivial insulator into a TI phase, then back to trivial insulator at
strong disorder. This is contrary to the general intuition that
disorder always tends to localize electronic states. This TAI phase
has since attracted extensive research interests\cite
{Jiang2009,Groth2009,Yamakage-11jpsj,Chen-11xxx,Xing-11prb,Li-11prb,HMGuo2010}.
In the original work, this phase was identified from the transport
properties showing a two-terminal conductance plateau $2e^{2}/h$
with extremely small fluctuations\cite{Li09}. Further numerical
studies confirmed that the plateau conductance in the TAI is
contributed from the dissipation-less edge states\cite{Jiang2009},
which further suggests the topological origin of this phenomena.
Theoretical study based on the first Born approximation of the
disordered Dirac fermions proposed that the TAI originates from a
band touching and subsequent re-opening of a topologically
nontrivial gap driven by disorder\cite{Groth2009}. The band touching
has been confirmed in the perturbative and numerical calculations,
but it is not sufficient to explain the whole region in the TAI.
Very recently, there has been a phase diagram for the disordered
HgTe/CdTe quantum spin Hall well, where the quantum spin Hall phase
and the TAI are connected\cite{Prodan2011}.

In this paper, we study the topological evolution of the TAI, and
examine the origin of the TAI from a topological point of view. We
calculate the band structure and the corresponding $Z_2$ invariants
as the disorder strength increases.  Starting with a topologically
trivial insulating phase, the bulk gap closes due to the disorder,
which changes the topological invariants of the occupied bands,
therefore triggering an insulator-TI transition (band inversion). As
the disorder strength further increases, a bulk gap is re-opened.
However, the gap value is too small (due to large sample size) and
too fluctuating (due to the randomness of disorder) to account for a
stable TAI phase as observed in transport calculations\cite{Li09}.
We shall show clear evidences that the TAI phase corresponds to a
continuous cluster of nontrivial subgaps, rather than a single gap.
These subgaps are separated by some extremely narrow subbands, and
survive through size scaling and random statistics. In other words,
in the TAI region, a Fermi level falls into a nontrivial subgap with
a probability close to 1, regardless of sample size and disorder
fluctuations. On the other hand, those extremely narrow subbands are
strongly localized, therefore they do not contribute to electronic
transport in the thermodynamic limit. This novel phase offers a new
realization of quantum spin Hall states in solids.

This paper is organized as follows. In section II, we describe the
model we use. In section III, the general definition and
calculation methods of topological invariants are reviewed. In
section IV, the ansatz of defining topological invariants for
disordered systems is introduced. The main results are described
in sections V and VI.

\section{II. The Model}

We first briefly revisit the Bloch's description of electronic properties.
In real space, the electronic Hamiltonian in a crystal lattice has the
general form
\begin{equation}
\mathcal{H}=\sum_i\sum_{\alpha\beta}\mathcal{H}_{\alpha,\beta}(i,i)c^{%
\dagger}_{i\alpha} c_{i\beta}+ \\
\sum_{\langle ij\rangle}\sum_{\alpha\beta}\mathcal{H}_{\alpha
\beta}(i,j)c^{\dagger}_{i\alpha} c_{j\beta},  \label{eqRealSpace}
\end{equation}
where $i,j$ are the indices of primary unit cells of the lattice, and $%
\alpha,\beta$ are the indices of freedom degree within the unit cell, e.g.,
sublattices, orbitals and spins etc. After Fourier transformation $%
c_{i\alpha}=\frac{1}{\sqrt{V}}\sum_{\bm{k}}c_{\bm{k}\alpha}e^{i\bm{k}\cdot{%
\bm{x}_i}}$, the Hamiltonian can be written as
\begin{equation}
\mathcal{H}=\sum_{\bm{k}}\sum_{\alpha\beta}H_{\alpha\beta}(\bm{k})c^{\dagger}_{\bm{k}\alpha}
c_{\bm{k}\beta},
\end{equation}
where $\bm{k}$ is defined in the first Brillouin zone. In the eigenproblem
\begin{equation}
\sum_{\beta}H_{\alpha\beta}(\bm{k})\; u_{n,\beta}(\bm{k})=E_{n}(\bm{k})\;
u_{n,\alpha}(\bm{k}),
\end{equation}
$E_{n}(\bm{k})$ determines the band structure, and $|u_{n}\rangle$ is the
unit cell periodic part of the Bloch function $|\psi_{n\bm{k}}\rangle=e^{i%
\bm{k}\cdot\bm{r}}|u_{n}(\bm{k})\rangle$.

The Bernevig-Hughes-Zhang (BHZ) model\cite{Bernevig2006}, a typical
tight-binding model with spin-orbit coupling that exhibits quantum spin Hall
phase, is defined on a square lattice with one $s$ orbital and one $p$
orbital on each site. In the above mentioned representation, the Bloch
Hamiltonian $H$ is a $4\times 4$ matrix written as
\begin{eqnarray}
H_{\alpha \beta }(\bm{k}) &=&\left(
\begin{array}{cc}
h(\bm{k}) & g(\bm{k}) \\
g^{\dagger }(\bm{k}) & h^{\ast }(-\bm{k})%
\end{array}%
\right)  \label{eqH} \\
h(\bm{k}) &=&d_{0}I_{2\times 2}+d_{1}\sigma _{x}+d_{2}\sigma
_{y}+d_{3}\sigma _{z}  \label{eqh} \\
g(\bm{k}) &=&\left(
\begin{array}{cc}
0 & -\Delta \\
\Delta & 0%
\end{array}%
\right)  \label{eqg} \\
d_{0}(\bm{k}) &=&-2D\big(2-\cos k_{x}-\cos k_{y}\big)  \notag \\
d_{1}(\bm{k}) &=&A\sin k_{x},\quad d_{2}(\bm{k})=-A\sin k_{y}  \notag \\
d_{3}(\bm{k}) &=&M-2B\big(2-\cos k_{x}-\cos k_{y}\big).  \notag
\end{eqnarray}%
Here $\alpha ,\beta $ are the indices of spinorbital within the unit cell, $%
\alpha ,\beta \in \{1,2,3,4\}\equiv \{|s\uparrow \rangle ,|p\uparrow \rangle
,|s\downarrow \rangle ,|p\downarrow \rangle \}$. $\sigma _{i}$ are Pauli
matrices acting on the spinor space spanned by $s$ and $p$ orbitals. The
real space Hamiltonian $\mathcal{H}$ of this model can be obtained from $%
H_{\alpha \beta }$ by a straightforward inverse Fourier transformation $c_{%
\bm{k}\alpha }=\frac{1}{\sqrt{V}}\sum_{\bm{i}}c_{i\alpha }e^{-i\bm{k}\cdot {%
\bm{x}_{i}}}$. The effect of non-magnetic impurities is included in real
space by adding a term
\begin{equation}
V_{I}=\sum_{i}\sum_{\alpha}U(i)c_{i\alpha }^{\dagger }c_{i\alpha },
\label{eqImpurity}
\end{equation}%
to $\mathcal{H}$, where $U(i)$ are random numbers uniformly distributed in $%
(-W/2,W/2)$.

\section{III. $Z_2$ Invariant}

\begin{figure}[htbp]
\includegraphics*[bb=10 0 540 550,width=0.45\textwidth]{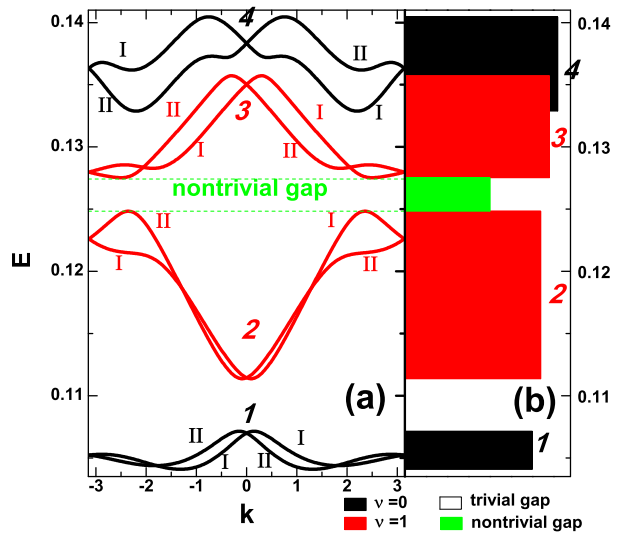}
\caption{(Color online) Schematic illustration of Kramers pairs.
Black is for trivial pair and red for non-trivial pair. (a) Band
 structures $E(\bm{k}%
) $. (b) Extension of the Kramers pair in (a), represented by the
width of a solid bar along the energy axis. Different heights (in
horizontal direction) of the bars are used to distinguish
individual pairs. Green bar is for nontrivial gap. }
\label{FSchematic}
\end{figure}
For a time reversal invariant system including both spin components,
Kramers Theorem states that, all the electronic bands $E_n(\bm{k})$
come in pairs connected at time reversal invariant points (TRIPs),
which are called Kramers pairs\cite{Fu2006,Fu2007}. If there are no
other degeneracies (e.g., disordered ``supercells'' which will be
discussed in the following) therefore each Kramers pair (KP) is
separated from others, a topological invariant can thus be defined
for each KP\cite{Essin2007}. In Fig. \ref{FSchematic} (a), we
illustrate the typical band structures of a time reversal invariant
system in the topological aspect. There are 8 bands, forming 4 KPs,
two of which (pairs 2 and 3 in red) are topologically nontrivial,
which will be defined below. Pairs 3 and 4 are separated but
overlapping in the energy axis. We will simply call the gap between
them as not fully gapped. Most of the information in Fig.
\ref{FSchematic} (a) can be plotted in a simple ``bar code'' version
in \ref{FSchematic} (b), where the extensions of the KPs and full
gaps along the energy axis are represented by the width of the bars
in this direction.

\begin{figure}[htbp]
\includegraphics*[width=0.38\textwidth]{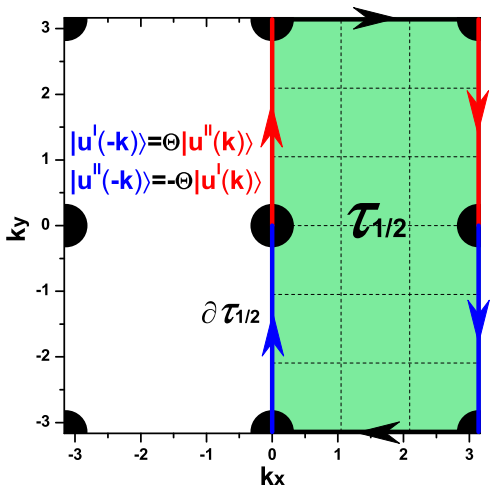}
\caption{(Color online) First Brillouin zone of a time reverse
symmetric solid in square lattice. Black dots are time reversal
invariant points. Green region is the effective Brillouin zone
$\protect\tau_{1/2}$, and arrows indicate its boundary
$\partial\protect\tau_{1/2}$. Dashed lines: mesh used in our
calculations (\protect\ref{eqZ2}).} \label{FBZ}
\end{figure}
In 2D the topological invariant $\nu$ associated with a KP is a
$Z_2$ integer defined from the periodic part of the Bloch function $u(\bm{k}%
) $ as\cite{Fu2006}
\begin{align}
\nu &=\frac{1}{2\pi}[\oint_{\partial\tau_{1/2}}d\bm{k}\cdot \bm{A}-
\int_{\tau_{1/2}}d\bm{k}^2 F]\mod 2,  \label{eqZ2} \\
\bm{A}(\bm{k})&=\sum_{s=\mathtt{I,II}}i\big\langle u^s(\bm{k})\big|%
\bm{\nabla}_{k}\big|u^s(\bm{k})\big\rangle,  \label{eqZ2b} \\
F(\bm{k})&=\Big(\nabla_{\bm{k}}\times \bm{A}(\bm{k})\Big)_z .  \label{eqZ2c}
\end{align}
Here $\tau_{1/2}$ is the effective Brillouin zone (EBZ) from which the rest
half can be obtained from its time reverse, $\partial\tau_{1/2}$ is the
boundary of $\tau_{1/2}$, as illustrated in Fig. \ref{FBZ}. The Romantic
numbers $\mathtt{I}$ and $\mathtt{II}$ in equation (\ref{eqZ2b}) label the
two branches of a KP (See Fig. \ref{FSchematic} (a)). Note the time reversal
constraint
\begin{equation}
\begin{aligned} |u^{\mathrm{I}}(-\bm{k})\rangle&=\Theta
|u^{\mathrm{II}}(\bm{k})\rangle\\ |u^{\mathrm{II}}(-\bm{k})\rangle&=-\Theta
|u^{\mathrm{I}}(\bm{k})\rangle \end{aligned}  \label{eqTRC}
\end{equation}
on $\partial\tau_{1/2}$ must be employed for equation (\ref{eqZ2}) to make
sense, where $\Theta=-is_y\otimes I_{2\times2} K$ is the time reversal
operator ($s_y$ is the Pauli matrix of physical spin and $K$ here is the
complex conjugation). A KP is trivial (nontrivial) if $\nu=0$ ($\nu=1$). The
topological invariant of a cluster of occupied KPs is just the sum of $\nu$
of all these KPs in the sense of $\mod 2$. Therefore, a gap between two
pairs are called trivial (nontrivial) if there are even (odd) number of
nontrivial pairs below it. If a gap is nontrivial, dissipationless edge
states will appear within the gap, when the system is truncated with open
boundaries\cite{Fu2006,Fu2007,Fu2007b}.

Among several equivalent definitions of $Z_2$ invariant $\nu$\cite
{Fu2007,Fu2006,Wang2010}, this definition has the advantages of
being expressed by well known topological quantities, i.e., Berry
connection $ \bm{A}$ and Berry curvature $\bm{F}$\cite{Xiao2010},
and being appropriate for numerical
evaluation\cite{Fukui2007,Essin2007,Xiao2010PRL}, which is briefly
introduced here. After discretizing the EBZ into a mesh (dashed
lines in Fig.\ref{FBZ} (c)), the field quantities $\bm{A}$ and
$\bm{F}$ can be defined from the eigenstates of the lattice
sites\cite{Fukui2007,Essin2007,Xiao2010PRL}, based on well-developed
lattice gauge theories. Note in the numerical calculations, for any
mesh site $\bm{k}$ in the EBZ, the phases of the eigenstates
(therefore the values of the field quantities) are arbitrarily and
independently determined by the numerical routines ($U(1)$ freedom
of local gauge choice). Care must be taken to cancel all these phase
uncertainties when summing up the discretized field quantities
$\bm{A}$ and $\bm{F}$ by equation (\ref{eqZ2}), so that the
resultant $\nu$ is gauge independent. Of course, the mesh should be
dense enough to obtain converged values for each KP.

For the clean systems, the topological properties of this model are
well understood\cite{Fu2007}, when the lower half bands are
occupied. When $ \Delta=0$, the quantum spin Hall phase with $Z_2=1$
is realized when $ 0<M/(2B)<2$. When tuning $M/B$, a ``band
inversion''\cite{Bernevig2006,Konig2008} occurs at $\Gamma$ point,
leading to a I-TI transition. The presence of $g(\bm{k})$ breaks the
conservation of $S_z$, but the topological invariants does not
change as long as the finite gap remains.

\section{IV. Zone Folding}

The topological invariants are defined in $\bm{k}$ space\cite%
{TKNN1982,Fu2006}, as introduced above. Impurities break the translation
invariance of the original lattice and make $\bm{k}$ badly defined. However,
for a disordered 2D sample with $N\times N$ unit cells, the above
topological arguments can be restored if twisted boundary conditions
\begin{equation}
\psi(\bm{r}+N\cdot \bm{a}_1)=e^{i\bm{k}\cdot N\bm{a}_1}\psi(\bm{r}),\quad
\psi(\bm{r}+N\cdot \bm{a}_2)=e^{i\bm{k}\cdot N\bm{a}_2}\psi(\bm{r})
\end{equation}
are introduced to the opposite boundaries of this finite sample,
where $\bm{a}_i$ are primitive vectors of the clean lattice\cite
{Niu1984,Sheng2006,Essin2007}. Physically speaking, this is
completely equivalent to taking this $N\times N$ sample as a large
unit cell of a 2D super-lattice, so that $\bm{k}$ can be defined in
a smaller Brillouin zone with reciprocal vectors $\bm{b}_i/N$, where
$\bm{b}_i$ is the reciprocal vector of the original lattice.
Disorder within this supercell tends to destroy all the band
degeneracies except those protected by time reversal symmetry, i.e.,
Kramers degeneracies. It is reasonable to imagine that for
sufficiently large $N$, the topological properties of this
superlattice can reflect those of the ``real'' disordered system. In
the following, we will call the primary unit cell of the original
clean system as a ``unit cell'', while the $N\times N$ sample as a
``supercell'' in this context.

In the clean limit, the band structure of the super-lattice can be derived
directly from that of the original lattice by using the standard method of
\textquotedblleft zone folding\textquotedblright , which is briefly reviewed
here. Now the Bloch Hamiltonian $H(\bm{k})$ becomes a $4N^{2}\times 4N^{2}$
matrix
\begin{equation}
H_{i\alpha ,\;j\beta }^{\mathrm{S}}(\bm{k}),\quad 1\leq i,j\leq N^{2},
\end{equation}%
where $\alpha ,\beta $ again represent the spinorbital indices within the
original unit cell, and $i,j$ are the indices of unit cells within the
supercell. The eigenvalues of $H^{\mathrm{S}}(\bm{k})$ are related with
those of the original lattice $E_{n}(\bm{k})$ as
\begin{equation}
E_{n,lm}^{\mathrm{S}}(\bm{k})=E_{n}(\bm{k}+\frac{l}{N}\bm{b}_{1}+\frac{m}{N}%
\bm{b}_{2}),\quad 0\leq l,m\leq N-1  \label{eqEigenvalues}
\end{equation}%
and associated eigenstates are
\begin{equation}
u_{n,lm}^{\mathrm{S}}(\bm{k})=\left(
\begin{array}{c}
e^{i(\frac{l}{N}\bm{b}_{1}+\frac{m}{N}\bm{b}_{2})\cdot \bm{r_1}}u_{n}(\bm{k}+%
\frac{l}{N}\bm{b}_{1}+\frac{m}{N}\bm{b}_{2}) \\
e^{i(\frac{l}{N}\bm{b}_{1}+\frac{m}{N}\bm{b}_{2})\cdot \bm{r_2}}u_{n}(\bm{k}+%
\frac{l}{N}\bm{b}_{1}+\frac{m}{N}\bm{b}_{2}) \\
\vdots \\
e^{i(\frac{l}{N}\bm{b}_{1}+\frac{m}{N}\bm{b}_{2})\cdot \bm{r_{N\times N}}%
}u_{n}(\bm{k}+\frac{l}{N}\bm{b}_{1}+\frac{m}{N}\bm{b}_{2})%
\end{array}%
\right) .  \label{eqEigenvectors}
\end{equation}%
Equations (\ref{eqEigenvalues}) and (\ref{eqEigenvectors}) can be verified
by a straightforward application of Bloch's Theorem, with the new
definitions of supercell and associated Brillouin zone in mind.

\section{V. A Small Supercell}

We will only consider the BHZ model in the case of $|D|<|B|$, so that the
system is always fully gaped between the lower and upper halves of bands,
when $M\neq0$. To obtain some insights from analytical treatments, we start
from a simple stage, a small supercell with $2\times2$ unit cells without
spin-flip parts, i.e., $\Delta=0$. Now the system is decoupled into two
sub-systems with single spin component, and the topological property of $%
H_{4\times4}(\bm{k})$ in equation (\ref{eqH}) can be reduced to that of $%
h_{2\times2}(\bm{k})$, represented by spin-resolved Chern
number\cite {Sheng2006,Prodan2010}. To further simplify the
analytical treatments, we can only consider the spin-up sub-system,
because its spin-down counterpart can be obtained from a
straightforward time reversal operation. For this
spin-up sub-system with a $2\times 2$-site supercell, the Hamiltonian is a $%
8\times8$ matrix
\begin{equation}
h_I^{\mathrm{S}}=h^{\mathrm{S}}(\bm{k})+V_I^{\mathrm{S}},  \label{eq4}
\end{equation}
where $h_{8\times8}^{\mathrm{S}}(\bm{k})$ is constructed from the above
zone-folding technique from the original $h_{2\times2}(\bm{k})$ in equation (%
\ref{eqh}) and the impurity term reads
\begin{equation}
V_I=W\cdot \mathrm{diag}(\epsilon_1,\epsilon_1,\epsilon_2,\epsilon_2,%
\epsilon_3,\epsilon_3,\epsilon_4,\epsilon_4),  \label{eqImpurity2}
\end{equation}
where $\epsilon_i$ are random numbers within the interval $(1/2,1/2)$ and $W$
is a single parameter to control the disorder strength. This $V_I$
represents random onsite potential distributed on 4 primary unit cells
within the $2\times2$ super-cell. We will show that, this minimal model in
equation (\ref{eq4}) that accommodates both disorder and topology, can
produce some non-trivial results.

Without impurities, as stated above, the eigenenergies and
eigenstates of $ h^{\mathrm{S}}$ can be constructed from those of
$h$ by zone-folding, eqs. (\ref{eqEigenvalues}) and
(\ref{eqEigenvectors}). The eigenenergies of $h^{\mathrm{S}}$ are
ordered by their values at the $\Gamma $ point $\bm{k}=0$ as
\begin{align*}
& E_{1,2,\cdots ,8}^{0}(\Gamma )=-8D-\big|M-8B\big|,\quad -4D-\big|M-4B\big|,
\\
& -4D-\big|M-4B\big|,\quad -M,\quad M,\quad -4D+\big|M-4B\big|, \\
& -4D+\big|M-4B\big|,\quad -8D+\big|M-8B\big|,
\end{align*}%
with a gap $2M$ between conductance band $E_{4}$ and valance band $E_{5}$.
The presence of $V_{I}$ will change band structures. Although the band
structures including impurities can be deduced from diagonalizing eq. (\ref%
{eq4}) directly, we will treat the impurities as a perturbation,
therefore the eigenenergies can be expressed as matrix elements
between unperturbed eigenstates $|u_{i}^{\mathrm{S}}\rangle $.
Straightforward calculations show that at $\Gamma =(0,0)$, the first
order correction to these two states is
\begin{eqnarray*}
E_{4}^{(1)}(\Gamma ) &=&\langle u_{4}^{S}(\Gamma )|V_{I}|u_{4}^{S}(\Gamma
)\rangle =\frac{W}{4}(\epsilon _{1}+\epsilon _{2}+\epsilon _{3}+\epsilon
_{4}) \\
E_{5}^{(1)}(\Gamma ) &=&\langle u_{5}^{S}(\Gamma )|V_{I}|u_{5}^{S}(\Gamma
)\rangle =\frac{W}{4}(\epsilon _{1}+\epsilon _{2}+\epsilon _{3}+\epsilon
_{4}),
\end{eqnarray*}%
which is just a uniform shift as a simple mean field of impurity potentials.
The second order correction is
\begin{eqnarray*}
E_{4}^{(2)}(\Gamma ) &=&\sum_{i\neq 4}\frac{\big|\langle u_{4}^{S}(\Gamma
)|V_{I}|u_{i}^{S}(\Gamma )\rangle \big|^{2}}{E_{4}(\Gamma )-E_{i}(\Gamma )}=%
\frac{-W^{2}F(\epsilon _{1},\epsilon _{2},\epsilon _{3},\epsilon _{4})}{%
128(B-D)} \\
E_{5}^{(2)}(\Gamma ) &=&\sum_{i\neq 5}\frac{\big|\langle u_{5}^{S}(\Gamma
)|V_{I}|u_{i}^{S}(\Gamma )\rangle \big|^{2}}{E_{5}(\Gamma )-E_{i}(\Gamma )}=%
\frac{W^{2}F(\epsilon _{1},\epsilon _{2},\epsilon _{3},\epsilon _{4})}{%
128(B+D)},
\end{eqnarray*}%
where
\begin{align*}
& F(\epsilon _{i})=5(\epsilon _{1}^{2}+\epsilon _{2}^{2}+\epsilon
_{3}^{2}+\epsilon _{4}^{2})-2(\epsilon _{1}\epsilon _{2}+\epsilon
_{2}\epsilon _{3}+\epsilon _{1}\epsilon _{4}+\epsilon _{3}\epsilon _{4}) \\
& -6(\epsilon _{1}\epsilon _{3}+\epsilon _{2}\epsilon _{4})%
\begin{cases}
=0,\quad \epsilon _{1}=\epsilon _{2}=\epsilon _{3}=\epsilon _{4} \\
>0,\quad \mathrm{otherwise}%
\end{cases}%
\end{align*}%
is a semi-positive-definite quadrics. Now the gap is
\begin{eqnarray}
E_{g} &=&\big|%
(E_{5}^{0}+E_{5}^{(1)}+E_{5}^{(2)})-(E_{4}^{0}+E_{4}^{(1)}+E_{4}^{(2)})\big|
\notag \\
&=&\bigg|2M\big(1+\frac{B}{M}\cdot \frac{W^{2}F(\epsilon _{1},\epsilon
_{2},\epsilon _{3},\epsilon _{4})}{64(B^{2}-D^{2})}\big)\bigg|.
\label{eqRenormalizedGap}
\end{eqnarray}

\begin{figure}[htbp]
\includegraphics*[width=0.42\textwidth]{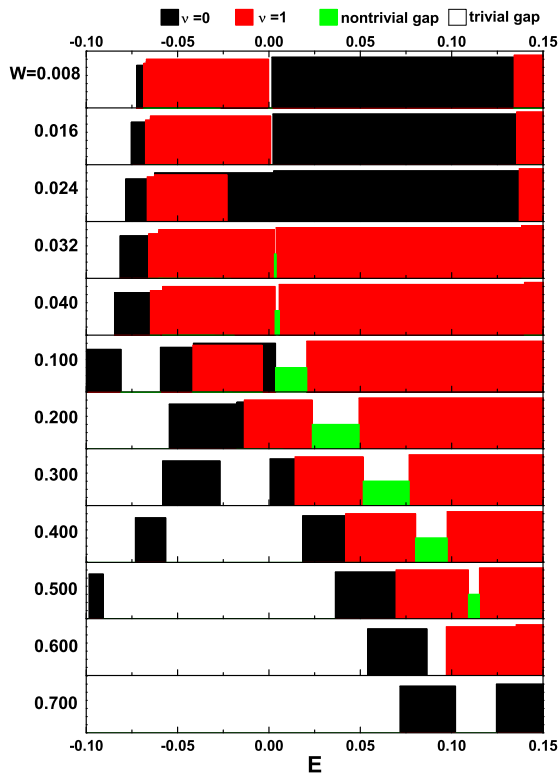}
\caption{Evolution of Kramers pairs and their topological invariants
for a $2\times2$ supercell as disorder strength $W$ increase (top to
bottom), for a definite configuration of $\{\epsilon_i\}$. Bars:
same as in Fig. \protect\ref{FSchematic}(b). Model parameters for H
in Eqs. (4): $A=0.0729$, $B=-0.0274$, $C=0$, $D=-0.205$, $M=0.001$
and $\Delta=0$, which are also consistent with
\protect\cite{Li09,Jiang2009}.  Lattice constant is set to be 1. }
\label{TwobyTwo}
\end{figure}
Equation (\ref{eqRenormalizedGap}) is the first important result in this
paper. It suggests that in the weak disorder regime, the impurities
effectively renormalize $M$\cite{Groth2009} and this renormalization comes
from a second order effect of disorder. Note the sign of this
renormalization term $\frac{B}{M}\cdot\frac{W^2 F(\epsilon_i)}{64(B^2-D^2)}$
does not depend on the concrete configuration of random impurities. If the
clean system is topologically non-trivial ($M/B>0$), the gap grows as $M+%
\mathrm{const}\cdot W^2$. This means that weak disorder tends to
make the two bands expel each other to avoid a band-touching which
will trigger a transition to a trivial insulator\cite{Hatsugai1999}.
This is a vivid manifestation of ``robustness against weak
disorder'' for TI. If the clean system is topologically trivial
($M/B<0$), on the other hand, the gap decays as
$M-\mathrm{const}\cdot W^2$. If $M$ is small, when we tune disorder
on, the gap will soon close at some small $W_c$, before strong
disorder makes the above perturbation treatment unreliable. This gap
close leads to an I-TI transition with the sign change of $M$. From
the topological point of view, the Chern number of $E_4$ changes by
1 after band-touching\cite
{Hatsugai1999,Murakami2007,Murakami2007b}. Remember we only
considered the spin-up block so far, but the physical argument of
topologically
trivial-nontrivial transition also applies when the time-reversal invariant $%
H$ involving both spin blocks is considered, by a simple correspondence
between Chern number and $Z_2$ invariant\cite{Moore2007}, as long as these
two blocks are decoupled.

To test the above physical pictures, we calculate the $Z_2$
topological invariants $\nu$ for a $2\times2$ supercell with both
spin components included. In Fig. \ref{TwobyTwo}, the evolution of
KPs for a definite configuration of $\{\epsilon _{i}\}$ with
increasing disorder strength $W$ are plotted. We can see that the
gap closing predicted by second order perturbation really happens at
$W=0.024$ and it does lead to a topological transition from $\nu=0$
to $\nu=1$ associated with the lower half bands. After this
transition, a topologically non-trivial gap (the green bar) emerges.
This gap will develop with further increasing $W$, until strong
disorder eventually close it again\cite{Hatsugai1999}. This
disorder-induced nontrivial gap shifts towards positive energy with
increasing $W$, as the TAI region observed in \cite{Li09} does. This
simple model itself also paves a route to producing a TI phase from
a trivial
insulator with spin-orbit coupling by constructing a superlattice\cite%
{ZFJiang2010}.

\section{VI. Large Supercells}

\begin{figure}[htbp]
\includegraphics*[width=0.5\textwidth]{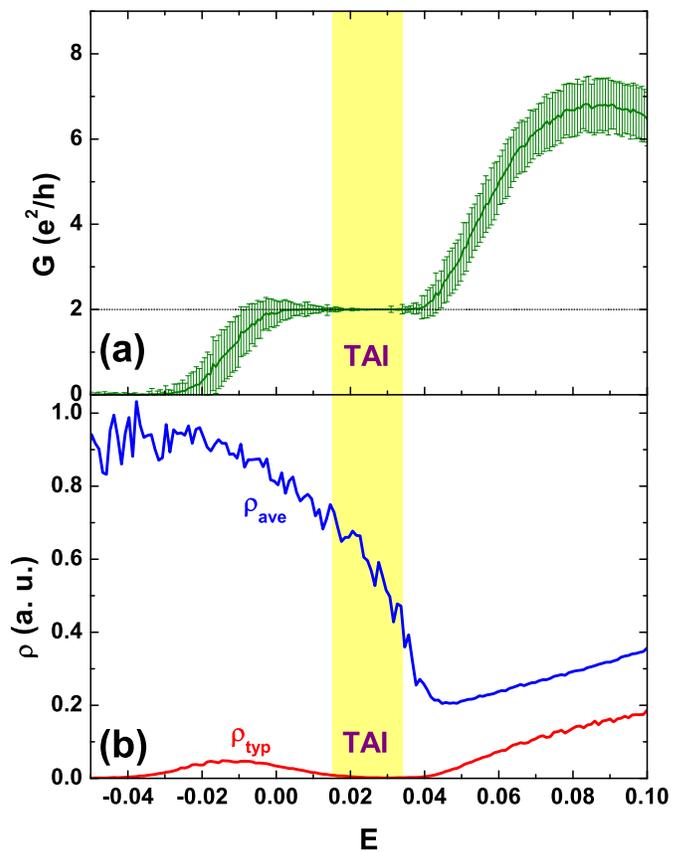}
\caption{(a) Two-terminal conductance of a samples with size $%
100\times100$, as functions of energy at a given disorder strength
$W=0.2$. Conductance is an average over 300 random configurations.
Conductance plateau with extremely small fluctuations corresponds
to the TAI phase, indicated by the light yellow region.
(b) The average DOS $\protect%
\rho_{\mathrm{ave}}$ (blue) and the typical DOS
$\protect\rho_{\mathrm{typ}}$ (red), calculated from 300 samples
with size $100\times 100$ and with periodic boundary conditions.
The parameters are the same as in Fig. 3} \label{FDOS}
\end{figure}

So far, the origin of TAI seems clear: the disorder triggers a band
touching, or a band inversion, after which a nontrivial gap opens for the
TAI phase to live. Unfortunately, this simple argument from a small
supercell cannot directly be applied to large samples, which will be shown
in the following. In Fig. \ref{FDOS} (a), we plot the statistics of
two-terminal conductance of a $100\times100$ samples. The two-terminal
conductance is calculated by the standard non-equilibrium Green's function
method\cite{Datta}, and the Fermi energy in the leads are fixed at $E_F^{%
\mathrm{lead}}=0.12$ to offer enough number of channels. The TAI phase is
identified as the conductance plateau $2e^2/h$ with extremely small
fluctuations, as in Ref. \cite{Li09}. If this region corresponds to a bulk
gap, the density of states (DOS) must vanish, at least in the case of
periodic boundary condition. The single particle local density of states
(LDOS) is calculated as\cite{MacKinnon1985}
\begin{equation}
\rho(i,E)=\frac{1}{N^2}\sum_n |\langle i|n\rangle|^2\delta(E-E_n).
\label{eqDOS}
\end{equation}
The arithmetic mean of the LDOS
\begin{equation}
\rho_{\mathrm{ave}}(E)\equiv \ll \rho(i,E) \gg
\end{equation}
is just the bulk DOS except for a constant factor, where $\ll \cdots \gg$ is
the arithmetic average over the sites of the sample. Meanwhile, the
geometric mean of the LDOS
\begin{equation}
\rho_{\mathrm{typ}}(E)\equiv \exp[ \ll \ln\rho(i,E) \gg]
\end{equation}
gives the localization property of the states. In the thermodynamic limit ($%
N\rightarrow\infty$), if $\rho_{\mathrm{ave}}(E)/\rho_{\mathrm{typ}%
}(E)\rightarrow0$, then the states around $E$ is localized\cite{Weisse2006}.
We thus plot $\rho_{\mathrm{ave}}$ and $\rho_{\mathrm{typ}}$ in Fig. \ref%
{FDOS} (b) in the case of periodic boundary conditions. Two remarkable
features can be read from the comparison between Fig. \ref{FDOS} (a) and
(b). First, the DOS $\rho_{\mathrm{ave}}$ does not vanish in TAI region. As
a matter of fact, there are regions with smaller DOS outside the TAI region.
In other words, the TAI phase does not live in a bulk gap at all. Second,
the TAI region corresponds to a vanishing of $\rho_{\mathrm{typ}}$. This
means that these states are extremely localized.

\begin{figure}[htbp]
\includegraphics*[width=0.42\textwidth]{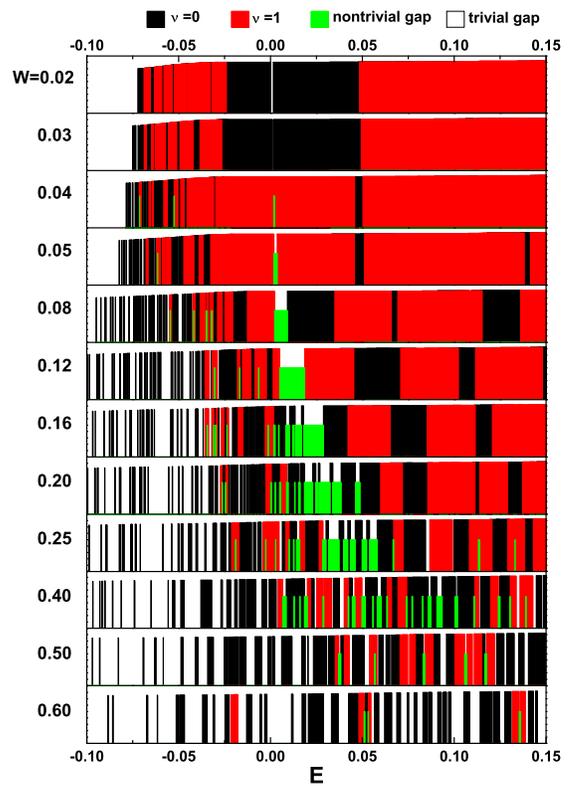}
\caption{Evolution of Kramers pairs and their topological invariants
for a $8\times8$ supercell, for a definite configuration of
$\{\epsilon_i\}$. Disorder strength $W$ increases from top to
bottom. Parameters are the same as in Fig. 3} \label{EightByEight}
\end{figure}
\begin{figure}[htbp]
\includegraphics*[bb=0 0 711 495,width=0.5\textwidth]{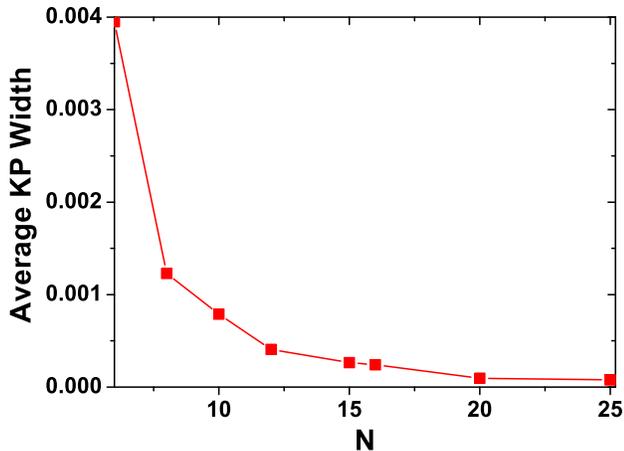}
\caption{(Color online) Average width of Kramers pairs between
$E=0.01$ and $E=0.03$ as a function of supercell size $N$.
Parameters are the same as in Fig. 3} \label{FigKPWidth}
\end{figure}

These surprising results throw a doubt on whether TAI can be
understood within the above mentioned topological regime. In order
to answer this, we repeat the numerical calculation of $\nu$ for
larger supercells. In Fig. \ref {EightByEight}, we plot the
evolution of KPs for a $8\times 8$ supercell associated with a
definite configuration of $\{\epsilon _{i}\}$, with increasing
disorder strength $W$. There is also a band touching at $W=0.05$,
which triggers a nontrivial subgap represented by a green bar, as in
the case of a small supercell. On the other hand, with stronger
disorder, for example, around $W=0.2$, it develops into a wider
region of nontrivial subgaps separated by narrow KPs, instead of one
single nontrivial gap. In Fig. \ref{FigKPWidth}, we show the average
width of KPs within the TAI region, it is clear that in the
thermodynamic limit, the KPs will be extremely narrow. These narrow
KPs are topologically trivial\cite{Tang2011,KSun2011,Neupert2011}
and do not affect the topology (trivial or nontrivial) of subgaps
between them. In other words, the disorder induced nontrivial nature
soon hides in the lower KPs deeply below the TAI region. We will
argue that, those narrow and topologically trivial KPs are
responsible for nonzero DOS in this region, while these nontrivial
subgaps are responsible for the TAI region observed from transport
calculations in Ref. \cite{Li09}.

\begin{figure}[tbph]
\includegraphics*[bb=0 0 711 495,width=0.5\textwidth]{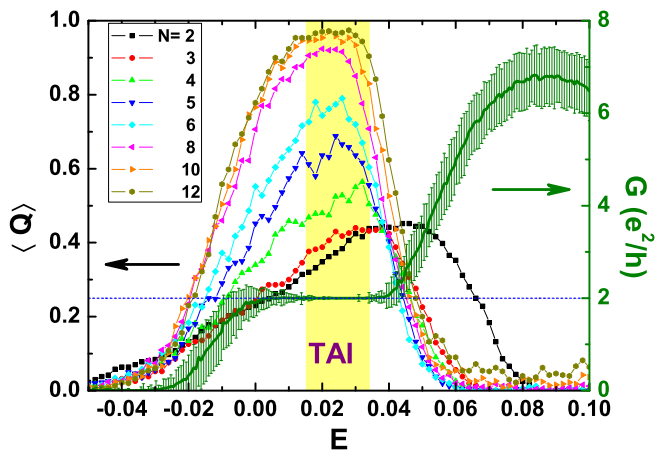}
\caption{(Color online) Probability of falling into a nontrivial subgap $%
\langle Q\rangle $ (lines with symbols) for supercells with
different size $N $, at $W=0.2$. Every $\langle Q\rangle $ is
averaged over 500 random configurations. Green curve is the
conductance, same as that in Fig. \protect\ref{FDOS}. Parameters
are the same as in Fig. 3} \label{FigD}
\end{figure}
The nonzero DOS is easy to understand. Although the KPs are
microscopically separated, but due to the broadening $\eta$
associated with any measure or calculation of DOS, they will give
rise to a continuous region of finite DOS. Moreover, these flat and
well separated KPs tend to be strong localized \cite{Essin2007}.
This is what we have observed in Fig. \ref{FDOS} (b). One may also
imagine that a fluctuating transition disorder strength $W_c$ from
sample to sample might contribute to the statistically nonzero DOS
in the TAI region. However, as stated in section V, since this gap
closing is a perturbation effect at weak disorder, the fluctuation
of $W_c$ is also very small. Indeed, the numerical results confirm
that (not shown here), compared to the width of TAI region, the
statistical error of $W_c$ is extremely small.

On the other hand, the origin of TAI phase exhibited from transport
calculations is more profound. It is well known that a nontrivial
subgap always gives rise to dissipationless edge states. However, in
our case, these disorder induced nontrivial subgaps are densely and
randomly distributed on the energy axis. To confirm that they are
indeed responsible for TAI phase, one must verify that these
nontrivial subgaps can survive through random statistics and size
scaling. To characterize this quantitatively, we define a function
\begin{equation}
Q(E)=\left\{ \begin{aligned} &1, \text{ $E\in$ a nontrivial and full subgap}
\\ &0, \text{ otherwise} \\ \end{aligned}\right.
\end{equation}%
for a definite supercell size and a definite disorder configuration.
The average $\langle Q(E)\rangle $ over the disorder ensemble is the
probability for the Fermi energy $E$ to fall into a nontrivial
subgap. If the above physical picture can make sense, $\langle
Q(E)\rangle $ must be a value very close to 1 in the TAI region,
under random averaging and size scaling. In Fig. \ref{FigD}, we show
the numerical results of $\langle Q(E)\rangle $ over averaging and
scaling. The conductance curve is also plotted as a comparison. It
is happy to see that, the $\langle Q(E)\rangle $ curves converge to
a broad peak approaching 1, and that the energy region of the broad
peak does correspond to the conductance plateau of TAI. Note this
broad peak of $\langle Q(E)\rangle $ calculated with supercell size
$\geq 10\times 10$ is sufficient to reproduce TAI region identified
from conductance plateau for $100\times 100$ samples. In the process
of scaling ($N\rightarrow \infty$), the numbers of subpairs and
subgaps increase, while the widths of individual subpairs and
subgaps decrease but with different decreasing rate. As a result, as
disclosed in Fig. \ref{FigD}, for large enough $N$, the total
measure of subgaps will dominate over that of flat subpairs,
$\langle Q\rangle\sim 1\gg1-\langle Q\rangle$. Fig. \ref{FigD} is
the most important result of this paper. It reveals that, the TAI
phase corresponds to a cluster of nontrivial subgaps instead of a
single topologically nontrivial gap. A Fermi energy falls into a
nontrivial subgap with a probability close to 1. The KPs, although
contributing to nonzero DOS in this region, are so narrow that their
measures on the energy axis are extremely small, and they are
localized therefore do not contribute to the electronic transport.
Because of the topological origin, it is now clear why the TAI is an
QSHE phase seen from the real space\cite{Prodan2011}, and why the
dissipationless currents are still carried by edge
states\cite{Jiang2009}.

\section{VII. Summary}

In summary, the topological evolution of TAI is studied in a supercell
regime. Starting from a trivial insulator phase with a small gap, weak
disorder inevitably lead to gap closing between the valence and conduction
bands, which is a second order perturbation effect. This causes an exchange
of topological invariants between them, and results in a transition to a
topologically nontrivial phase. In the limit of large supercell, there will
be very large numbers of subbands and subgaps densely distributed on the
energy axis. However, there exists a continuous region where the Fermi
energy falls into a nontrivial subgap with an extremely high probability,
even after a statistical average over the disorder ensemble. This special
region can thus support a stable and observable TAI phase. This physical
picture also helps find disorder-induced topological insulators in other
materials and higher dimensions\cite{HMGuo2010}.

\section{Acknowledgements}

This work was supported by the Resarch Grant Council of Hong Kong
under Grant No. HKU705110P.


\begin{thebibliography}{99}
\bibitem{Hasan2010} M. Z. Hasan and C. L. Kane, Rev. Mod. Phys. \textbf{82},
3045 (2010).

\bibitem{Qi2010} X. L. Qi and S. C. Zhang, Phys. Today 63, 33 (2010)

\bibitem{Moore2010} J. E. Moore, Nature 464, 194 (2010)

\bibitem{Kane2005A} C. L. Kane and E. J. Mele, Phys. Rev. Lett. \textbf{95},
146802 (2005).

\bibitem{Fu2007} L. Fu and C. L. Kane, Phys. Rev. B \textbf{76}, 045302
(2007).

\bibitem{Moore2007} J. E. Moore and L. Balents, Phys. Rev. B \textbf{75},
121306(R) (2007).

\bibitem{Qi2008} X.-L. Qi, T. L. Hughes and S.-C. Zhang, Phys. Rev. B
\textbf{78}, 195424 (2008).

\bibitem{Kane2005B} C. L. Kane and E. J. Mele, Phys. Rev. Lett. \textbf{95},
226801 (2005).

\bibitem{Bernevig2006} A. Bernevig, T. Hughes and S. C. Zhang, Science
\textbf{314}, 1757 (2006).

\bibitem{Li09} J. Li, R.-L. Chu, J. K. Jain and S.-Q. Shen, Phys. Rev. Lett.
\textbf{102}, 136806 (2009).

\bibitem{Hatsugai1999} Y. Hatsugai, K. Ishibashi and Y. Morita, Phys. Rev.
Lett. \textbf{83}, 2246 (1999).

\bibitem{Konig-06science} M. Konig, S. Wiedmann, C. Brune, A. Roth, H.
Buhmann, L. W. Molenkamp, X. L. Qi and S. C. Zhang, Science 318, 766 (2007).

\bibitem{Konig2008} M. K\"{o}nig, H. Buhmann, L. W. Molenkamp, T. Hughes,
C.-X. Liu, X.-L. Qi and S.-C. Zhang, J. Phys. Soc. Jpn. \textbf{77}, 031007
(2008).

\bibitem{Jiang2009} H. Jiang, L. Wang, Q.-F. Sun and X. C. Xie, Phys. Rev. B
\textbf{80}, 165316 (2009).

\bibitem{Groth2009} C. W. Groth, M. Wimmer, A. R. Akhmerov, J. Tworzyd{\l }o
and C. W. J. Beenakker, Phys. Rev. Lett. \textbf{103}, 196805 (2009).

\bibitem{Yamakage-11jpsj} A. Yamakage, K. Nomura, K. Imura, Y. Kuramoto, J.
Phys. Soc. Jpn. 80 053703 (2011)

\bibitem{Chen-11xxx} L. Chen, Q. Liu, X. Lin, X. G. Zhang, and X. Y. Jiang,
arXiv: 1106.4103

\bibitem{Xing-11prb} Y. X. Xing, L. Zhang, and J. Wang, Phys. Rev. B 84,
035110 (2011)

\bibitem{Li-11prb} W. Li, J. D. Zang, and Y. J. Jiang, Phys. Rev. B 84,
033409 (2011)

\bibitem{HMGuo2010} H.-M. Guo, G. Rosenberg, G. Refael and M. Franz, Phys.
Rev. Lett. \textbf{105}, 216601 (2010).

\bibitem{Prodan2011} E. Prodan, Phys. Rev. B \textbf{83}, 195119 (2011).

\bibitem{Fu2006} L. Fu and C. L. Kane, Phys. Rev. B \textbf{74}, 195312
(2006).

\bibitem{Essin2007} A. M. Essin and J. E. Moore, Phys. Rev. B \textbf{76},
165307 (2007).

\bibitem{Fu2007b} L. Fu, C. L. Kane and E. J. Mele, Phys. Rev. Lett. \textbf{%
98}, 106803 (2007).

\bibitem{Wang2010} Z. Wang, X.-L. Qi and S.-C. Zhang, New J. Phys. \textbf{12%
}, 065007 (2010).

\bibitem{Xiao2010} D. Xiao, M.-C. Chang and Q. Niu, Rev. Mod. Phys. \textbf{%
82}, 1959 (2010).

\bibitem{Fukui2007} T. Fukui and Y. Hatsugai, J. Phys. Soc. Jpn. \textbf{76}%
, 053702 (2007).

\bibitem{Xiao2010PRL} D. Xiao, Y. G. Yao, W. X. Feng, J, Wen, W. G. Zhu,
X.-Q. Chen, G. M. Stocks and Z. Y. Zhang, Phys. Rev. Lett. \textbf{105},
096404 (2010).

\bibitem{TKNN1982} D. J. Thouless, M. Kohmoto, M. P. Nightingale and M. den
Nijs, Phys. Rev. Lett. \textbf{49}, 405 (1982).

\bibitem{Niu1984} Q. Niu and D. J. Thouless, J. Phys. A \textbf{17}, 2453
(1984).

\bibitem{Sheng2006} D. N. Sheng, Z. Y. Weng, L. Sheng and F. D. M. Haldane,
Phys. Rev. Lett. \textbf{97}, 036808 (2006).

\bibitem{Prodan2010} E. Prodan, T. L. Hughes and B. A. Bernevig, Phys. Rev.
Lett. \textbf{105}, 115501 (2010).

\bibitem{Murakami2007} S. Murakami, S. Iso, Y. Avishai, M. Onoda and N.
Nagaosa, Phys. Rev. B \textbf{76}, 205304 (2007).

\bibitem{Murakami2007b} S. Murakami, New J. Phys. \textbf{9}, 256 (2007).

\bibitem{ZFJiang2010} Z.-F. Jiang, R.-L. Chu and S.-Q. Shen, Phys. Rev. B
\textbf{81}, 115322 (2010).

\bibitem{Datta} S. Datta, \emph{Electronic Transport in Mesoscopic Systems}
(Canmbridge University Press, Cambridge, U.K., 1995).

\bibitem{MacKinnon1985} A. MacKinnon, Z. Phys. B - Condensed Matter \textbf{%
59}, 385 (1985).

\bibitem{Weisse2006} A. Weisse, G. Wellein, A. Alvermann and H. Fehske, Rev.
Mod. Phys. \textbf{78}, 275 (2006).

\bibitem{Tang2011} E. Tang, J.-W. Mei and X.-G. Wen, Phys. Rev.
Lett. \textbf{106}, 236802 (2011).

\bibitem{KSun2011} K. Sun, Z.-C. Gu, H. Katsura and S. Das Sarma, Phys. Rev.
Lett. \textbf{106}, 236803 (2011).

\bibitem{Neupert2011} T. Neupert. L. Santos, C. Chamon and C. Mudry, Phys. Rev.
Lett. \textbf{106}, 236804 (2011).
\end{thebibliography}
\end{document}